\documentclass{aa}

\usepackage{graphicx}
\usepackage{txfonts}
\usepackage{color}
\usepackage{amstext}

\begin{document}

   \title{Stellar populations in the Carina region.}
   \subtitle{The Galactic plane at $l = 291^{\circ}$}
   \titlerunning{Stellar populations in Carina}

   \author{Molina-Lera J.A\inst{1},
           Baume G.\inst{1,2},
           Gamen R.\inst{1,2},
           Costa E.\inst{3},
          \and
           Carraro G.\inst{4,5}
          }
   \authorrunning{Molina Lera et al.}

   \institute{Instituto de Astrof\'{\i}sica de La Plata (CONICET, UNLP),
              Paseo del Bosque s/n, La Plata, Argentina \\
              \email{jalejoml@fcaglp.unlp.edu.ar (AML)}
   \and
              Facultad de Ciencias Astron\'omicas y Geof\'{\i}sicas (UNLP),
              Paseo del Bosque S/N, La Plata, (B1900FWA), Argentina \\
              \email{gbaume@fcaglp.unlp.edu.ar (GB)} \\
              \email{rgamen@fcaglp.unlp.edu.ar (RG)}
   \and
              Departamento de Astronom\'{\i}a, Universidad de Chile,
              Casilla 36-D, Santiago, Chile \\
              \email{costa@das.uchile.cl (EC)}
   \and
              ESO, Alonso de Cordova 3107, Vitacura, Santiago de Chile, Chile \\
              \email{gcarraro@eso.org (GC)}
   \and
              Dipartimento di Fisica e Astronomia, Universitá di Padova, Italy\\
             }

   \date{Received September **, ****; accepted March **, ****}


  \abstract
  {Previous studies of the Carina region have revealed its complexity and richness as well as a significant number of early-type stars. However, in many cases, these studies only concentrated on the central region (Trumpler~14/16) or were not homogeneous. This latter aspect, in particular, is crucial because very different ages and distances for key clusters have been claimed in recent years.}
  { The aim of this work is to study in detail an area of the Galactic plane in Carina, eastward $\eta$ Carina. We analyze the properties of different stellar populations and focus on a sample of open clusters and their population of YSOs and highly reddened early stars. We also studied the stellar mass distribution in these clusters and the possible scenario of their formation. Finally, we outline the Galactic spiral structure in this direction.}
  {We obtained deep and homogeneous photometric data ($UBVI_{KC}$) for six young open clusters: NGC~3752, Trumpler~18, NGC~3590, Hogg~10, 11, and 12,  located in Carina at $l \sim 291^{\circ}$, and their adjacent stellar fields, which we complemented  with spectroscopic observations of a few selected targets. We also culled  additional information from the literature, which includes stellar spectral classifications and near-infrared photometry from $2MASS$. We finally developed a numerical code that allowed us to perform a homogeneous and systematic analysis of the data. Our results provide more reliable estimates of distances, color excesses, masses, and ages of the stellar populations in this direction.}
  {We estimate the basic parameters of the studied clusters and find that they identify  two overdensities of young stellar populations located at about 1.8~kpc and 2.8~kpc, with $E_{B-V} \sim 0.1-0.6$. We find evidence of pre-main-sequence populations inside them, with an apparent coeval stellar formation in the most conspicuous clusters. We also discuss apparent age and distance gradients in the direction NW-SE. We study the mass distributions of the covered clusters and several others in the region (which we took form the literature). They consistently show a canonical  IMF slope (the Salpeter one). We discover and characterise an abnormally reddened massive stellar population, scattered between 6.6 and 11~kpc. Spectroscopic observations of ten stars of this latter population show that all selected targets were massive OB stars. Their location is consistent with the position of the Carina-Sagittarius spiral arm.}
  {}


   \maketitle
%

\section{Introduction} \label{sec:intro}

Young stellar clusters and associations are ideal tracers of the spiral structure of the Milky Way and, more in general, of the global thin-disk structure, defined by its radial and vertical scale length. Insights on the disk warp and flare can also be obtained by their spatial distribution. All these are crucial quantities that in turn allow us to compare the Milky Way to external spirals (\citealt{2009A&A...499L..21G}), with the ultimate aim of understanding which type of spiral our galaxy is and how it was assembled (\citealt{2009PASP..121..213C} and \citealt{2013MNRAS.433.2986W}).

To unravel the three-dimensional structure of the disk, we need precise distance estimates, which rely on accurate photometry and spectroscopy and on a proper assessment of the interstellar absorption. This is a complicated property of the intervening interstellar medium and in our Galaxy is known to vary from direction to direction (\citealt{2012Ap&SS.337..303T}).

Young stellar clusters are also invaluable probes of the star formation process when a good handling of their ages is possible. Their spatial location together with their age distribution can help us assess whether star formation in a given Galactic region was coeval, sequential, or triggered (\citealt{2011A&A...531A..73B}).

Finally, the mass distribution of stars within young stellar clusters is by definition very close to the initial mass function (IMF). Variations of the mass function slope from one region of the Galaxy to the other can highlight possible environmental effects during the early evolution of a star cluster, such as mass loss induced by shocks or massive stellar winds, and stellar escapes caused by close encounters (\citealt{2013MNRAS.433.1378L}).

To properly address all these fundamental questions, we require the accumulation of high quality data for several objects in a given Galactic sector. A recent outstanding example is the delineation of the spiral structure in the Galactic anticenter
in both the second and third Galactic quadrant (\citealt{2015A&A...577A.142M} and \citealt{2008ApJ...672..930V}), together with the characterization of the young stellar populations of  the Galactic warp and flare (\citealt{2015AJ....149...12C}) and the discussion of the disk radial extent (\citealt{2010ApJ...718..683C}).

One of the most massive star-forming regions in the Milky Way is the great Carina nebula (\citealt{2011A&A...530A..34P},
\citealt{2012Ap&SS.337..303T}) and the region surrounding it in the Galactic plane ($\sim 283^{\circ} < l < \sim 292^{\circ}$). Its complexity and richness makes it an intriguing target to investigate the topics we have outlined above. This region is characterized by many young stellar clusters (\citealt{1995RMxAC...2...57F}) and a large number of early-type stars ($\sim$~60~O-type near the Carina nebula alone; \citealt{2006MNRAS.367..763S}). This has been interpreted as evidence that the line of sight to Carina crosses the Carina-Sagittarius Galactic spiral arm (\citealt{1970IAUS...38..246B}, \citealt{1976AJ.....81..155C}, \citealt{1976PASP...88..647H}, \citealt{2000A&A...357..308G}, \citealt{2012PASP..124..128K}, \citealt{2009A&A...493...71C}, \citealt{2010PASP..122..516P} and references therein).

In spite of the many studies, the precise location and nature of the Carina-Sagittarius arm is far from being settled. Very different distances for the same clusters (see, e.g., \citealt{2004A&A...418..525C}; \citealt{2013A&A...555A..50C}) have been reported by different groups. In addition, there is a lack of consensus whether the Carina-Sagittarius arm is a primary or a secondary spiral arm. Optical studies favor the idea that it is a grand design arm (\citealt{2003A&A...397..133R}), while radio studies support the idea it is a secondary structure (\citealt{2009PASP..121..213C}).

The star formation in Carina has been investigated in detail near the Carina nebulae (\citealt{2011A&A...530A..34P}), and evidence has been brought forth of an age gradient (\citealt{2004A&A...418..525C}), which would support a sequential mode of star formation. Extending the coverage of the region outside the classical central zone would be the next step to better characterize this fascinating Galactic sector. With this goal in mind, we carried out deep photometric observations covering the so far unexplored Galactic plane near $l = 291^{\circ}$ .

\begin{figure}
\begin{center}
\includegraphics[width=8.5cm]{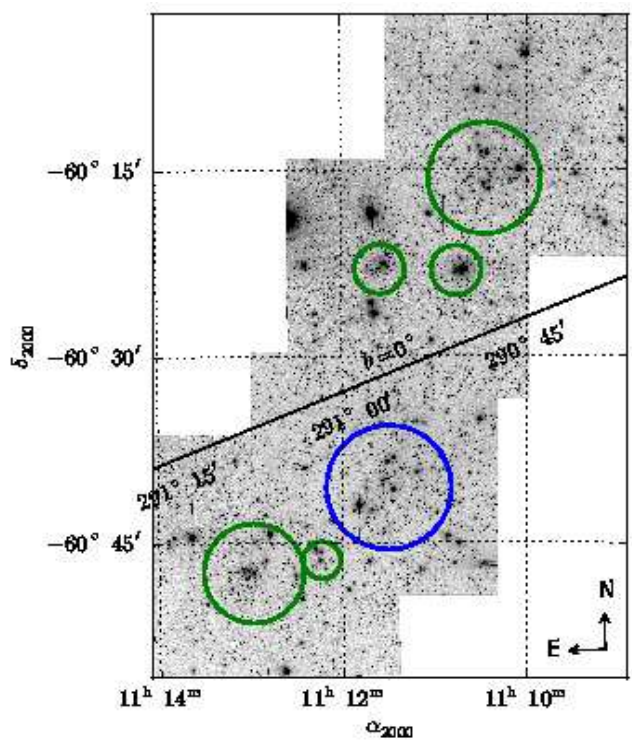}
\caption{Mosaic of the area covered in this study built from our CCD $V$ band images. The central solid line is the Galactic plane ($b = 0^{\circ}$). Circles indicate the adopted location and sizes of the studied clusters, and their colors represent the corresponding stellar populations (see Sect.~\ref{sec:populations}).}
\label{fig:ccds}
\end{center}
\end{figure}

Our survey includes three large young open clusters (NGC~3572, Trumpler~18, and NGC~3590), three minor ones (Hogg~10, 11, and 12), and a large part of their surrounding fields (see Fig.~\ref{fig:ccds}). We complemented this data set with spectroscopic observations of a handful of peculiar objects in the area and spectroscopic stellar classifications available in the literature.

The layout of the paper is as follows: in Sect.~\ref{sec:data} we describe our data set and the basic reduction procedures; in Sect.~\ref{sec:analysis} we present the method used to analyze the data, in Sect.~\ref{sec:clusters} we derive the main properties of the clusters, and in Sect.~\ref{sec:populations} we discuss the different stellar populations found and their connection with the Galactic structure. Finally, we summarize our results in Sect. 6.


\section{Data} \label{sec:data}

\subsection{$UBVI_{KC}$ photometry} \label{sec:phot_data}

\subsubsection{Observations}

The photometry was secured in March 2006 and March 2009 with the Y4KCAM camera onboard the Cerro Tololo Inter American Observatory (CTIO, Chile) 1.0 m telescope, which was operated by the SMARTS\footnote{http://www.astro.yale.edu/smarts} consortium. This camera is equipped with an STA $4064\times4064$ CCD with $15 \mu$ pixels, which provides a field of view (FOV) of $20\farcm0 \times 20\farcm0$ and a scale of $0\farcs289$/pix. Other detector characteristics can be found at the Y4Cam web-page\footnote{http://www.astronomy.ohio-state.edu/Y4KCam/detector.html}.
This configuration allowed us to cover the targeted clusters entirely and at the same time to sample a significant portion of the Galactic field around each of them (see Fig.~\ref{fig:ccds}).

In our 2006 run we covered the regions of NGC~3590/Hogg~12 (Frame~1), Hogg~10~and~11 (Frame~2), and NGC~3572 (Frame~3), and in 2009 that of Trumpler~18 (Frame~4). All observations were carried out in photometric conditions. A detailed description of the observations is given in Table~\ref{tab:frames}.


\begin{table*}
\begin{center}
\fontsize{9} {12pt}
\selectfont
\caption{Description of the observations.}
\label{tab:frames}
\begin{tabular}{ccccc@{\extracolsep{7pt}}ccccc}
\hline
Frame & \multicolumn{2}{c}{Date} &
 \multicolumn{4}{c}{Center} & \multicolumn{2}{c}{Airmass} & $FWHM$ \\
\cline{4-7}
& & & \multicolumn{2}{c}{$\alpha_{J2000}$} & \multicolumn{2}{c}{$\delta_{J2000}$} & & & \\
\hline
 1 & \multicolumn{2}{c}{03/19/06}    & \multicolumn{2}{c}{11:12:44.5} &
     \multicolumn{2}{c}{-60:46:07.4} & \multicolumn{2}{c}{1.17-1.23}  & $1\farcs0-1\farcs4$ \\
 2 & \multicolumn{2}{c}{03/20/06}    & \multicolumn{2}{c}{11:11:16.6} &
     \multicolumn{2}{c}{-60:23:53.8} & \multicolumn{2}{c}{1.17-1.26}  & $1\farcs0-1\farcs4$ \\
 3 & \multicolumn{2}{c}{03/22/06}    & \multicolumn{2}{c}{11:10:14.9} &
     \multicolumn{2}{c}{-60:12:13.6} & \multicolumn{2}{c}{1.17-1.24}  & $1\farcs2-1\farcs5$ \\
 4 & \multicolumn{2}{c}{03/20/09}    & \multicolumn{2}{c}{11:11:39.3} &
     \multicolumn{2}{c}{-60:39:32.1} & \multicolumn{2}{c}{1.16-1.32}  & $0\farcs9-1\farcs4$ \\
\hline
\hline
\multicolumn{10}{c}{Exposure times [$s$]} \\
\hline
& \multicolumn{4}{c}{Run 2006 (Frames 1,2,3)} & \multicolumn{4}{c}{Run 2009 (Frame 4)} & \\
\cline{2-5} \cline{6-9}
 & $U$  & $B$  & $V$ & $I_{KC}$ & $U$  & $B$  & $V$ & $I_{KC}$ & \\
\hline
 & 1800 &  900 & 700 &      600 & 2000 & 1500 & 900 &      900 & \\
 &  200 &  100 & 100 &      100 &  200 &  150 & 100 &      100 & \\
 &   30 &   30 &  30 &       30 &   30 &   20 &  10 &       30 & \\
 &   10 &    7 &   5 &        5 &    - &    - &   - &        - & \\
\hline
\end{tabular}
\end{center}
\end{table*}


\begin{table*}
\begin{center}
\fontsize{9} {14pt}\selectfont
\caption{Derived photometric calibration coefficients for each frame/night.}
\label{tab:coef}
\begin{tabular}{cllll}
\hline
 Coef. &
 \multicolumn{1}{c} {19/03/06} &
 \multicolumn{1}{c} {20/03/06} &
 \multicolumn{1}{c} {22/03/06} &
 \multicolumn{1}{c} {20/03/09} \\
\hline
 $u_{1}$   & $+3.258\pm0.007$ & $+3.261\pm0.007$ & $+3.270\pm0.008$ & $+3.121\pm0.007$ \\
 $u_{2}$   & $+0.45$          & $+0.45$          & $+0.45$          & $+0.45$          \\
 $u_{3}$   & $-0.022\pm0.015$ & $-0.009\pm0.018$ & $-0.017\pm0.014$ & $-0.023\pm0.010$ \\
 $rms$     & $+0.03$          & $+0.04$          & $+0.04$          & $+0.05$          \\
\hline
 $b_{1}$   & $+2.031\pm0.015$ & $+2.015\pm0.018$ & $+2.019\pm0.013$ & $+1.932\pm0.010$ \\
 $b_{2}$   & $+0.25$          & $+0.25$          & $+0.25$          & $+0.25$          \\
 $b_{3}$   & $+0.138\pm0.022$ & $+0.152\pm0.025$ & $+0.170\pm0.016$ & $+0.132\pm0.010$ \\
 $rms$     & $+0.03$          & $+0.04$          & $+0.03$          & $+0.05$          \\
\hline
 $v_{1vb}$ & $+1.820\pm0.012$ & $+1.807\pm0.021$ & $+1.821\pm0.010$ & $+1.873\pm0.006$ \\
 $v_{2vb}$ & $+0.16$          & $+0.16$          & $+0.16$          & $+0.16$          \\
 $v_{3vb}$ & $-0.080\pm0.015$ & $-0.058\pm0.027$ & $-0.055\pm0.012$ & $-0.036\pm0.006$ \\
 $rms$     & $+0.02$          & $+0.04$          & $+0.02$          & $+0.03$          \\
\hline
 $v_{1vi}$ & $+1.808\pm0.009$ & $+1.809\pm0.018$ & $+1.826\pm0.010$ & $+1.876\pm0.006$ \\
 $v_{2vi}$ & $+0.16$          & $+0.16$          & $+0.16$          & $+0.16$          \\
 $v_{3vi}$ & $-0.059\pm0.010$ & $-0.054\pm0.020$ & $-0.052\pm0.011$ & $-0.032\pm0.005$ \\
 $rms$     & $+0.02$          & $+0.04$          & $+0.02$          & $+0.04$          \\
\hline
 $i_{1}$   & $+2.703\pm0.016$ & $+2.672\pm0.018$ & $+2.709\pm0.014$ & $+2.687\pm0.005$ \\
 $i_{2}$   & $+0.08$          & $+0.08$          & $+0.08$          & $+0.08$          \\
 $i_{3}$   & $-0.048\pm0.017$ & $+0.003\pm0.023$ & $-0.003\pm0.015$ & $-0.014\pm0.003$ \\
 $rms$     & $+0.03$          & $+0.03$          & $+0.03$          & $+0.04$          \\
\hline
\end{tabular}
\end{center}
\end{table*}


\subsubsection{Reduction and calibration} \label{sec:phot_red}

Science frames were pre-processed in a standard way following the guidelines given in \cite{2011A&A...531A..73B} and using the {\sc iraf}\footnote{IRAF is distributed by NOAO, which is operated by AURA under cooperative agreement with the NSF.}
package {\sc ccdred}. For this purpose zero exposures and sky dome flats were taken every night. The photometry was performed with the {\sc iraf daophot} and {\sc photcal} packages. Instrumental magnitudes were derived using the point spread function (PSF) method (\citealt{1987PASP...99..191S}), using a quadratic spatially variable PSF. About 20-30  homogeneously distribute bright stars in each image were selected to construct the PSF models and to estimate the corresponding aperture corrections. Finally,  the data from different frames were combined for
each filter using the code {\sc daomaster}
\citep{1992ASPC...25..297S}.

To determine the transformation equations between our instrumental system to the standard $UBVI_{KC}$ system as well as the nightly extinction correction,  we observed standard stars every night ($\sim30$ per night) from the catalog of \cite{ 1992AJ....104..340L} (fields PG~1323-086, SA~101, SA~104 and SA~107), with an air-mass range between 1.1 and 2.1. These standard star fields allowed for a wide variety of colors of the standards.  Aperture photometry was carried out for all the standard stars using the {\sc iraf photcal} package. To tie our photometry to the standard system, the following transformation equations were used:

\begin{center}
\begin{tabular}{lc}
$u = U + u_1     + u_2     \cdot (U-B)      + u_3     \cdot X$ &  \\
$b = B + b_1     + b_2     \cdot (B-V)      + b_3     \cdot X$ &  \\
$v = V + v_{1vi} + v_{2vi} \cdot (V-I_{KC}) + v_{3vi} \cdot X$ &  \\
$v = V + v_{1vi} + v_{2vi} \cdot (V-I_{KC}) + v_{3vi} \cdot X$ &  \\
$i = I_{c} + i_1 + i_2     \cdot (V-I_{KC}) + i_3     \cdot X,$ &  \\
\end{tabular}
\end{center}

\noindent where $UBVI_{KC}$ and $ubvi$ are the standard and instrumental magnitudes, respectively, and $X$ is the airmass of the observation. Table~\ref{tab:coef} summarizes the transformation and extinction coefficients obtained for each night. Finally, the data from the four nights were combined, again using {\sc daomaster}.

To estimate the photometric completeness of our data, we carried out several artificial-star experiments on our long exposure frames (see \citealt{2005A&A...436..527C}). For this process, we added artificial stars (preserving the same color and luminosity distributions as that of our real populations) in random positions to our images by means of the {\sc iraf} task {\sc addstar}, and repeated the reduction procedure described above. To keep crowding of the artificial frames at a level
similar to that of our true images, the amount of artificial stars added was only 10\% of the real stellar population in each frame.

Completeness factors ($CF$) for different $V$ magnitude bins were computed for each experiment. They are defined as the ratio between the number of artificial stars recovered and the number of stars added. In Table~\ref{tab:complet} we present the results of our experiments. The $CF$ shown are average values for each night. $N$ is the number of real stars that were detected in each magnitude bin.


\begin{table*}
\begin{center}
\caption{Completeness factors as a function of $V$ for each
  frame/night.}
\label{tab:complet}
\fontsize{9} {14pt}\selectfont
\begin{tabular}{lcrrrrrrr}
\hline
        & $V~bin$   & 15-16 & 16-17 & 17-18  & 18-19 & 20-21 & 21-22 & 22-23 \\
\hline
Frame~1 & $N$       &   561 &  1183 &   2066 &  3411 &  4766 &  4926 &  1811 \\
        & $CF~[\%]$ & 100.0 &  94.6 &   93.2 &  89.4 &  81.5 &  36.2 &   3.8 \\
\hline
Frame~2 & $N$       &   687 &  1358 &   2559 &  4097 &  5019 &  3573 &   309 \\
        & $CF~[\%]$ & 100.0 &  94.2 &   90.6 &  85.5 &  70.2 &  14.5 &   2.5 \\
\hline
Frame~3 & $N$       &   600 &  1115 &   2172 &  3505 &  4333 &  3089 &   115 \\
        & $CF~[\%]$ & 100.0 &  95.2 &   93.7 &  88.9 &  82.9 &  41.6 &   1.6 \\
\hline
Frame~4 & $N$       &   605 &  1290 &   2313 &  3674 &  5182 &  5807 &  3689 \\
        & $CF~[\%]$ & 100.0 &  94.5 &   93.2 &  89.4 &  83.4 &  70.1 &  30.0 \\
\hline
\end{tabular}
\end{center}
\end{table*}


\subsection{Spectroscopic data} \label{sec:spec_data}

\subsubsection{Observations} \label{sec:spec_obs}

We performed follow-up  spectroscopic observations of the brightest ten objects of a sample of stars that we name {\it Population C} (see Sect.~\ref{sec:populations}), based on the photometric analysis (see Sect.~\ref{sec:analysis}). We employed the GMOS \footnote{https://www.gemini.edu/sciops/instruments/gmos/}
spectrograph in its long-slit mode attached to the 8 m telescope of Gemini South (Chile), using "poor-weather" time\footnote{proposal GS-2015A-Q-101; 8.4h; PI: R. Gamen}. We employed a slit of $1\farcs0$ width at a central wavelength of 5200 \AA\, and 5300 \AA\, to avoid the gap between CCDs, and the grating B600. We obtained spectra with a typical resolution $R \sim 1000$. Suitable exposure times were estimated by means of the Integration Time Calculator (ITC) provided by the Gemini consortium, considering several combinations of poor-weather conditions. We finally decided to take two 900~s exposures for each star.

\subsubsection{Reductions and spectral classification} \label{sec:spec_class}

Spectra were processed in a standard way\footnote{http://www.gemini.edu/sciops/data/\\~~IRAFdoc/gmos\textunderscore longslit\textunderscore example.cl},
using the {\sc gemini} package at {\sc iraf}. Because of changing weather conditions, the signal-to-noise ratio ($S/N$) of our spectra varies significantly, but all of them are suitable to identify early-type stars (see Sect.~\ref{sec:spec_class}).

Stellar spectral classification is quite a standard and solid process, which relies on the identification of spectral features (emission or absorption lines) in an observed spectrum, and the comparison of it  with well-established class (spectral and luminosity) prototypes (see, e.g., Jashek \& Jashek 1990). Therefore, the
spectral classification of our targets was made by visual comparison with standard spectra of OB stars, which in our case were taken from the suite of spectral types reported by Sota et al. (2011).

The analysis was carried out using the code {\sc mgb} by \citealt{2015hsa8.conf..603M}. This code allows simultaneously displaying the standard spectrum and the one that is to be classified. Classification is then performed  by adopting the spectral type and luminosity class of the standard template that better reproduces the features of the observed spectrum.

\begin{figure}
\begin{center}
\includegraphics[width=9cm]{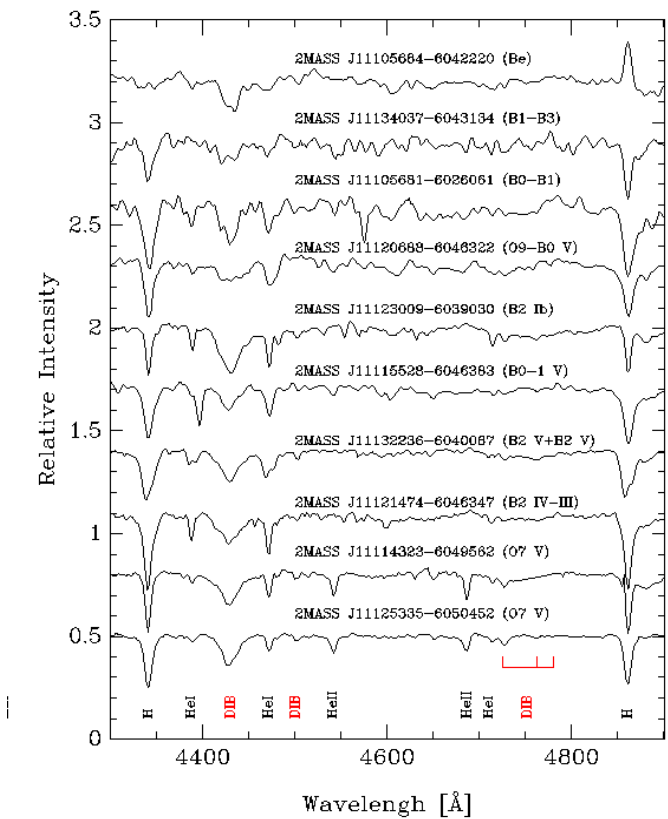}
\caption{Spectra of the ten OB-type stars observed for this work. Most relevant spectral lines are indicated. See text for details on their classification.}
\label{fig:st_gemini}
\end{center}
\end{figure}

Naturally, the process involves some degree of visual inspection and some subjectivity when features cannot be
clearly distinguished. This occurs often if the spectrum $S/N$ is poor. When the identification of  characteristic spectral features was not obvious, we determined the spectral types by adopting different tailored criteria.
In Fig.~\ref{fig:st_gemini} we present the spectra of the ten program stars (two O-type and eight  B-type stars), while in the following we provide details on their individual classification.

\begin{itemize}
\item 2MASS J11105681-6026061 (\#1035): this source has a low $S/N$ ($<30$) spectrum. We were able to identify many He {\sc i} absorption lines, however,
such as $\lambda$4388, $\lambda$4471, $\lambda$4921, and $\lambda$5875, and, marginally, He {\sc ii} $\lambda$4541. We were unable to detect Mg {\sc ii} $\lambda$4481. We therefore classified this star as B0-B1. Based on the broadening of the Balmer lines and because no metallic lines were clearly detected, we assigned a luminosity class V.
\item 2MASS J11105684-6042220 (\#519): the spectrum shows H$\alpha$ and H$\beta$ in emission, and H$\epsilon$ in absorption. H$\gamma$ and H$\delta$ are too marginal to be of use. We also identified emission lines of He {\sc i} and Fe {\sc ii}. No conspicuous absorption lines could be identified to determine the spectral type, therefore  we classified this star as Be.
\item 2MASS J11114323-6049562 (\#767): we clearly identified He {\sc ii} in its spectrum. Comparison with standard spectra indicates an O7 V spectral type.
\item 2MASS J11115528-6046383 (\#920): another source with a
low $S/N$ ($<$30) spectrum. Most lines are quite noisy. We were
able to identify some He {\sc i} absorption lines, however, such as $\lambda$4471, $\lambda$4921, $\lambda$5015, and $\lambda$5875, and also He {\sc ii} $\lambda$4686. We classified this star as B0-B1 V.
\item 2MASS J11120688-6046322 (\#810): a very noisy spectrum. We were able to identify absorption lines of He {\sc ii} $\lambda$4542, $\lambda$4686, and $\lambda$5411, and other He {\sc i} lines. We therefore assigned an O9-B0~V spectral type.
\item 2MASS J11121474-6046347 (\#1204): comparison with standard stellar spectra indicates a B2 spectral type. Considering the broadening of the Balmer lines, the luminosity class may be IV or III.
\item 2MASS J11123009-6039030 (\#508): comparison with standard stellar spectra indicates a B2 Ib spectral type.
\item 2MASS J11125335-6050452 (\#900): comparison with standard stellar spectra indicates a O7 V spectral type.
\item 2MASS J11132236-6040067 (\#1185): the spectrum shows double lines of He {\sc i,} and a similar feature can be noted in the Balmer lines. The relative intensities are quite similar, suggesting that the observed spectrum is the superposition of those of two B1-B2 stars, probably forming a binary system. We adopted this classification because Mg {\sc ii} was detected, ruling out spectral types later than B3 (where this line was as intense as other He {\sc i} lines visible in our spectrum).
\item 2MASS J11134037-6043134 (\#978): this was one of the poorest spectra obtained, with a $S/N$ lower than 20 in the spectral range used for classification. We were able to identify He {\sc i} $\lambda$5875 in absorption and weak Balmer lines, however. Considering that the former line becomes intense at approximately B2, we classified this star as B1-B3.
\end{itemize}



\subsection{Complementary data} \label{sec:otherdata}

We have compared our $UBVI_{KC}$ photometry with previous photometric measurements available for our areas of interest, compiled from the WEBDA\footnote{http://www.univie.ac.at/webda} database and
the APASS\footnote{https://www.aavso.org/apass} catalog. In this latter case we transformed the $gri$ data to the $VI_C$ system using the transformations given by \cite{2005AJ....130..873J}. The mean differences obtained for the two data sets are shown in Table~\ref{tab:deltas}. The above bibliographic sources were also used to obtain photometric data (which were shifted to our photometric
system) for a few bright stars that are saturated in our images. This helped in the analysis of our photometric diagrams (see Sect.~\ref{sec:upperms}).

The Two-Micron All Sky Survey catalog (2MASS; \citealt{2003yCat.2246....0C}, \citealt{2006AJ....131.1163S}) was used to carry out an astrometric calibration using the procedure indicated in \cite{2009MNRAS.398..221B}, which led to a positional precision of $\sim 0\farcs17$ in both coordinates.

Finally, using the $SIMBAD$ database and results reported
by \cite{2009yCat.1280....0K}, we collected existing spectral classification information in the covered region.


\begin{table*}
\begin{center}
\caption{Mean differences between our $UBVI_{KC}$ data and previous photometric measurements, in the sense ours minus others ($\Delta$).}
\label{tab:deltas}
\fontsize{9} {14pt}\selectfont
\begin{tabular}{lccccccc}
\hline
 & APASS           &
 NGC~3572$^{1}$    &
 Hogg~10$^{2}$     &
 Hogg~11$^{1}$     &
 Trumpler~18$^{3}$ &
 Hogg~12$^{1}$     &
 NGC~3590$^{2}$    \\
\hline
 $\Delta U$ & --    & -0.02 & +0.01 & -0.03 & -0.02 & -0.02 & -0.00 \\
 $\Delta B$ & -0.04 & -0.05 & +0.00 & -0.13 & -0.09 & -0.07 & -0.05 \\
 $\Delta V$ & -0.04 & -0.05 & +0.05 & -0.01 & -0.07 & -0.11 & -0.06 \\
 $\Delta I$ & -0.04 & --    & --    & --    & -0.11 & --    & --    \\
 $N$        & 210   & 8     & 8     & 5     & 24    & 5     & 21    \\
\hline
\end{tabular}
\fontsize{9} {12pt}\selectfont
\begin{minipage}{13cm}
\vspace{0.2cm}
{\bf Note:} Photoelectric measurements adopted as references: \\
1 = \cite{1975A&AS...20..125M}; 2 = \cite{1976AJ.....81..155C}; 3 = \cite{1990A&AS...86..209V}.
\end{minipage}
\end{center}
\end{table*}


\subsection{Final catalog} \label{sec:catalogue}

We used the {\sc STILTS\footnote{http://www.star.bris.ac.uk/~mbt/stilts/}} tool to manipulate tables and cross-correlate our $UBVI_{KC}$ data with the 2MASS photometry and with the spectral classification available in the literature or performed by us.  The correlation procedure between optical data and 2MASS data was made using $1\farcs0$ as matching radius. This method only matches the nearest sources and does not attempt to deblend sources in cases of high crowding.

We produced then a catalog with astrometric and photometric ($UBVI_{KC}$+$JHK$) information for about $63000$ objects in the region around $l \sim 291^{\circ}$ shown in Fig.~\ref{fig:ccds}, together with the spectroscopic classification for several stars included in Table~\ref{tab:spectra}. The full catalog is made available in electronic form at the Centre de Donnais Stellaire (CDS) website.

\section{Method} \label{sec:analysis}

We analyzed the data set using a numerical code developed in {\sc fortran} and {\sc gnuplot}. This code allowed us to homogeneously and systematically study the data. Therefore, we were able to reduce potential bias of the adopted parameters for the clusters and stellar populations.

\subsection{Spectrophotometric study} \label{sec:spec}


\begin{table*}
\caption{Main parameters for stars with available spectral classification ($SC$).}
\label{tab:spectra}
\fontsize{9}{10pt}
\selectfont
\centering
\begin{tabular}{lcclcccccc}
\hline
 ID & $\alpha_{J2000.0}$ & $\delta_{J2000.0}$ & $SC$ & $V$ & $(B-V)$ & $E_{U-B}$  & $E_{B-V}$ &  $E_{V-I}$ & $V_{\circ}-M_V$ \\
\hline
\multicolumn{9}{l} {NGC~3572} \\
\hline
 HD~97166              & 11:10:06.0 & -60:14:54.3 & O7.5 IV + O9 III & ~7.89~ & ~0.03~ & -          & -         & -         &  -    \\
 ~~~HD~97166a          & 11:10:06.0 & -60:14:54.3 & O9 III           & ~8.50* & ~0.04* & ~0.27      & 0.35      & -         &  12.5 \\
 ~~~HD~97166b          & 11:10:06.0 & -60:14:54.3 & O7.5 IV          & ~8.80* & ~0.02* & ~0.27      & 0.35      & -         &  12.5 \\
\hline
\multicolumn{9}{l} {Hogg~10} \\
 HD~97253              & 11:10:42.1 & -60:23:04.2 & O5 IIIe          & ~7.11~ & ~0.18~ & ~0.44      & 0.50      & -         & 11.3 \\
 174                   & 11:10:46.7 & -60:23:00.7 & B5 IV            & 11.86~ & ~0.28~ & ~0.26      & 0.45      & 0.56      & -    \\
 260                   & 11:10:53.0 & -60:22:24.4 & B3 V             & 12.32~ & ~0.24~ & ~0.34      & 0.45      & 0.59      & 12.6 \\
 336                   & 11:10:38.9 & -60:22:30.1 & B5 V             & 12.71~ & ~0.26~ & ~0.31      & 0.43      & 0.56      & 12.4 \\
 357                   & 11:10:37.9 & -60:24:00.4 & B5 V             & 12.77~ & ~0.25~ & ~0.25      & 0.42      & 0.51      & 12.5 \\
 385                   & 11:10:39.7 & -60:23:10.4 & B2.5 V           & 12.85~ & ~0.26~ & ~0.46      & 0.48      & 0.63      & 13.4 \\
 557                   & 11:10:39.3 & -60:22:04.2 & B6 V             & 13.36~ & ~0.29~ & ~0.31      & 0.44      & 0.59      & 12.7 \\
 949                   & 11:10:38.2 & -60:23:03.8 & B6 V             & 14.01~ & ~0.36~ & ~0.69      & 0.51      & 0.61      & 13.1 \\
\hline
\multicolumn{9}{l} {Hogg~11} \\
\hline
 HD~97381              & 11:11:32.7 & -60:22:38.2 & B1 III           & ~8.32~ & ~0.02~ & ~0.25      & 0.29      & 1.26      & 11.6 \\
\hline
\multicolumn{9}{l} {Trumpler~18} \\
\hline
 HD~97434              & 11:11:49.6 & -60:41:58.3 & O7.5 III         & ~8.06~ & ~0.08~ & ~0.43      & 0.39      & -         & 12.3 \\
 120                   & 11:10:59.6 & -60:43:50.0 & B7 V             & 11.54~ & ~0.18~ & ~0.19      & 0.31      & 0.51      & 11.0 \\
\hline
\multicolumn{9}{l} {NGC~3590} \\
\hline
 110                   & 11:13:03.1 & -60:47:52.0 & B2 V             & 11.40~ & ~0.32~ & ~0.45      & 0.56      & 0.71      & 12.2 \\
 171                   & 11:13:01.1 & -60:47:22.1 & B4 V             & 11.83~ & ~0.33~ & ~0.33      & 0.52      & 0.64      & 11.5 \\
 231                   & 11:12:57.4 & -60:47:36.3 & B5 V             & 12.19~ & ~0.38~ & ~0.37      & 0.55      & 0.68      & 11.5 \\
 251                   & 11:13:09.4 & -60:47:18.0 & B5 V             & 12.28~ & ~0.36~ & ~0.35      & 0.52      & 0.67      & 11.7 \\
 306                   & 11:12:59.2 & -60:47:34.4 & B5 V             & 12.54~ & ~0.36~ & ~0.43      & 0.53      & 0.66      & 11.9 \\
 339                   & 11:12:57.1 & -60:47:00.9 & B4 V             & 12.72~ & ~0.36~ & ~0.34      & 0.55      & 0.66      & 12.3 \\
 343                   & 11:12:52.3 & -60:46:32.3 & B5 V             & 12.73~ & ~0.36~ & ~0.34      & 0.52      & 0.65      & 12.1 \\
 412                   & 11:13:06.2 & -60:47:01.0 & B5 V             & 12.98~ & ~0.39~ & ~0.67      & 0.56      & 0.65      & 12.2 \\
 547                   & 11:12:58.9 & -60:46:36.3 & B6 V             & 13.34~ & ~0.34~ & ~0.39      & 0.49      & 0.59      & 12.5 \\
 596                   & 11:12:51.3 & -60:46:58.0 & B9 V             & 13.42~ & ~0.47~ & ~0.17      & 0.54      & 0.70      & 11.5 \\
 669                   & 11:13:03.3 & -60:46:57.9 & B6 V             & 13.55~ & ~0.33~ & ~0.36      & 0.48      & 0.61      & 12.8 \\
 813                   & 11:13:13.7 & -60:47:18.3 & B8 V             & 13.80~ & ~0.45~ & ~0.40      & 0.56      & 0.72      & 12.2 \\
 1018                  & 11:13:00.9 & -60:46:40.6 & B9 V             & 14.10~ & ~0.38~ & ~0.29      & 0.45      & 0.53      & 12.4 \\
 1171                  & 11:13:13.5 & -60:47:54.4 & B9 V             & 14.27~ & ~0.50~ & ~0.35      & 0.57      & 0.75      & 12.2 \\
\hline
\multicolumn{9}{l} {Population~$C$} \\
\hline
 767                   & 11:11:43.2 & -60:49:56.3 & O7 V             & 13.72~ & ~0.92~ & ~1.07      & 1.25      & 1.86      & 14.3 \\
 900                   & 11:12:53.4 & -60:50:45.0 & O7 V             & 13.94~ & ~1.24~ & ~1.35      & 1.57      & 2.38      & 13.3 \\
 508                   & 11:12:30.1 & -60:39:03.0 & B2 Ib            & 13.26~ & ~1.44~ & ~1.33      & 1.60      & 2.09      & 14.2 \\
 519                   & 11:10:56.9 & -60:42:22.3 & Be               & 13.28~ & ~0.86~ & -          & -         & -         & -    \\
 810                   & 11:12:06.9 & -60:46:32.3 & O9 V - B0 V      & 13.80~ & ~0.91~ & ~1.08-1.03 & 1.23-1.22 & 1.79-1.79 & 13.9-13.4 \\
 920                   & 11:11:55.3 & -60:46:38.3 & B0 V - B1 V      & 13.96~ & ~1.03~ & ~1.07-0.97 & 1.34-1.31 & 1.99-1.81 & 13.1-12.6 \\
 978                   & 11:13:40.2 & -60:43:13.5 & B1 V - B3 V      & 14.05~ & ~1.10~ & ~1.05-0.82 & 1.37-1.31 & 1.90-1.81 & 12.5-11.2 \\
 1035                  & 11:10:56.8 & -60:26:06.1 & B0 V - B1 V      & 14.13~ & ~1.00~ & ~1.00-0.90 & 1.31-1.28 & 2.08-1.90 & 13.4-12.9 \\
 1185                  & 11:13:22.4 & -60:40:06.8 & B2 V + B2 V      & 14.28~ & ~0.87~ & -          & -         & -         & -    \\
 ~~~1185a              & 11:13:22.4 & -60:40:06.8 & B2 V             & 15.03* & ~0.87* & ~0.68      & 1.11      & 2.17      & 13.6 \\
 ~~~1185b              & 11:13:22.4 & -60:40:06.8 & B2 V             & 15.03* & ~0.87* & ~0.68      & 1.11      & 2.17      & 13.6 \\
 1204                  & 11:12:14.8 & -60:46:34.7 & B2 III - B2 IV   & 14.30~ & ~0.86~ & ~0.89-0.92 & 1.09-1.10 & 1.51-1.51 & 14.1 \\
\hline
\multicolumn{9}{l} {Field} \\
\hline
 HD~97223              & 11:10:31.3 & -60:35:05.8 & B3 V             & ~8.73~ & -0.04~ & ~0.38      & 0.17      & -         & ~9.9 \\
 HD~97222              & 11:10:32.4 & -60:08:38.5 & B0 II            & ~8.84~ & ~0.20~ & ~0.42      & 0.49      & -         & -    \\
 HD~97284              & 11:10:52.3 & -60:44:30.0 & B0.5 Iab-Ib      & ~8.86~ & ~0.05~ & ~0.36      & 0.28      & 0.36      & 14.5 \\
 HD~97297              & 11:11:04.0 & -60:18:34.2 & A0 IV            & ~8.90~ & ~0.19~ & -0.04      & 0.20      & 0.15      & -    \\
 CPD-593120            & 11:10:47.2 & -60:23:19.1 & B5 II - III      & 10.22~ & ~0.33~ & ~0.34      & 0.50      & -         & 10.8 \\
\hline
\end{tabular}
\begin{minipage}{17cm}
\vspace{0.2cm} {\bf Note:} (*) Value computed from the individual spectral classification of the binary component.
\end{minipage}
\end{table*}


Our spectroscopic observations, together with spectroscopic data available in the literature, allowed us to estimate the main parameters (distances and color excesses) of the 41 stars in the region of interest. To derive these parameters, we applied the traditional spectrophotometric method and used the absolute magnitude $M_V$ and intrinsic color calibrations provided by \cite{2013JKAS...46..103S} and \cite{1978Obs....98...54C}. Linear interpolations were applied for targets whose spectral type was not included in these latter works. Depending on the stellar group, values of $R_V = A_V/E_{B-V} = 3.1$ or $3.5$ were assumed (see Sects.~\ref{sec:upperms} and \ref{sec:populations}).

For bona-fide binary systems with a double spectral classification (stars HD~97166 and \#1185), the individual magnitudes were first derived and then the previous procedure was applied to each individual member. For stars with luminosity classes (LC) IV and II, we adopted intrinsic colors from LC V and III, respectively. For stars with spectral types earlier than B0.5V, we adopted $(V-I)_{\circ} = -0.47$. The resulting parameters are presented in Table~\ref{tab:spectra}.

\subsection{Cluster parameters} \label{sec:parameters}


\begin{table*}
\caption{Fundamental parameters of the studied open clusters.}
\label{tab:clusters}
\fontsize{9} {14pt}
\selectfont
\centering
\begin{tabular}{llllcccccc}    
\hline
\multicolumn{1}{c}{Cluster}    &
\multicolumn{1}{c}{Work}       &
\multicolumn{2}{c}{Center}     &
\multicolumn{1}{c}{Radius}     &
 $E_{(B-V)}$                   &
 $V_{\circ}-M_V$               &
\multicolumn{2}{c}{Age [Myrs]} &
 $\Gamma$                      \\
\cline{3-4}\cline{8-9}
 &&
 $\alpha_{J2000.0}$ & $\delta_{J2000.0}$ &
 $[']$              &
 $[mag]$            &
 $[mag]$            &
 Nuclear            &
 Contraction        &
                    \\
\hline
 NGC~3572     & This work & 11:10:27.4 & -60:15:40.2 & 4.5 & 0.36 & 11.9 & $\sim3$  & $\sim5$    & $-1.07 \pm 0.24$ \\
              & D2002     & 11:10:23   & -60:14:54   & 5.0 & 0.39 & 11.5 & $7.8$    & -          & -                \\
\hline
 Hogg~10      & This work & 11:10:45.7 & -60:23:05.1 & 2.0 & 0.46 & 12.0 & $\sim3$  & -          & -                \\
              & D2002     & 11:10:42   & -60:24:00   & 3.0 & 0.46 & 12.2 & $6.0$    & -          & -                \\
\hline
 Hogg~11      & This work & 11:11:35.3 & -60:22:58.9 & 2.0 & 0.35 & 12.0 & $\sim10$ & -          & -                \\
              & D2002     & 11:11:37   & -60:24:00   & 2.0 & 0.32 & 11.8 & $12.0$   & -          & -                \\
\hline
 Trumpler~18  & This work & 11:11:29.8 & -60:40:34.1 & 5.0 & 0.21 & 10.8 & $\sim25$ & $\sim20$ & $-1.33 \pm 0.34$   \\
              & D2002     & 11:11:28   & -60:40:00   & 5.0 & 0.32 & 10.7 & $15.5$   & -          & -                \\
\hline
 Hogg~12      & This work & 11:12:13.8 & -60:46:23.9 & 1.5 & 0.46 & 11.6 & $\sim10$ & -          & -                \\
              & D2002     & 11:13:01   & -60:47:00   & 4.0 & 0.40 & 11.5 & $30.2$   & -          & -                \\
\hline
 NGC~3590     & This work & 11:12:58.4 & -60:47:25.0 & 4.0 & 0.45 & 11.7 & $\sim25$ & $\sim10^*$ & $-1.54 \pm 0.71$ \\
              & D2002     & 11:12:59   & -60:47:18   & 3.0 & 0.45 & 11.1 & $17.0$   & -          & -                \\
\hline
\end{tabular}
\begin{minipage}{18cm}
\vspace{0.2cm}
\hspace{2cm} {\bf Note:} (*) See Sect.~\ref{sec:ngc3590} for the interpretation of the estimated contraction age.
D2002 = \cite{2002A&A...389..871D}.
\end{minipage}
\end{table*}


\subsubsection{Centers and sizes of the clusters} \label{sec:sizes}


\begin{figure*}
\begin{center}
\includegraphics[width=18cm]{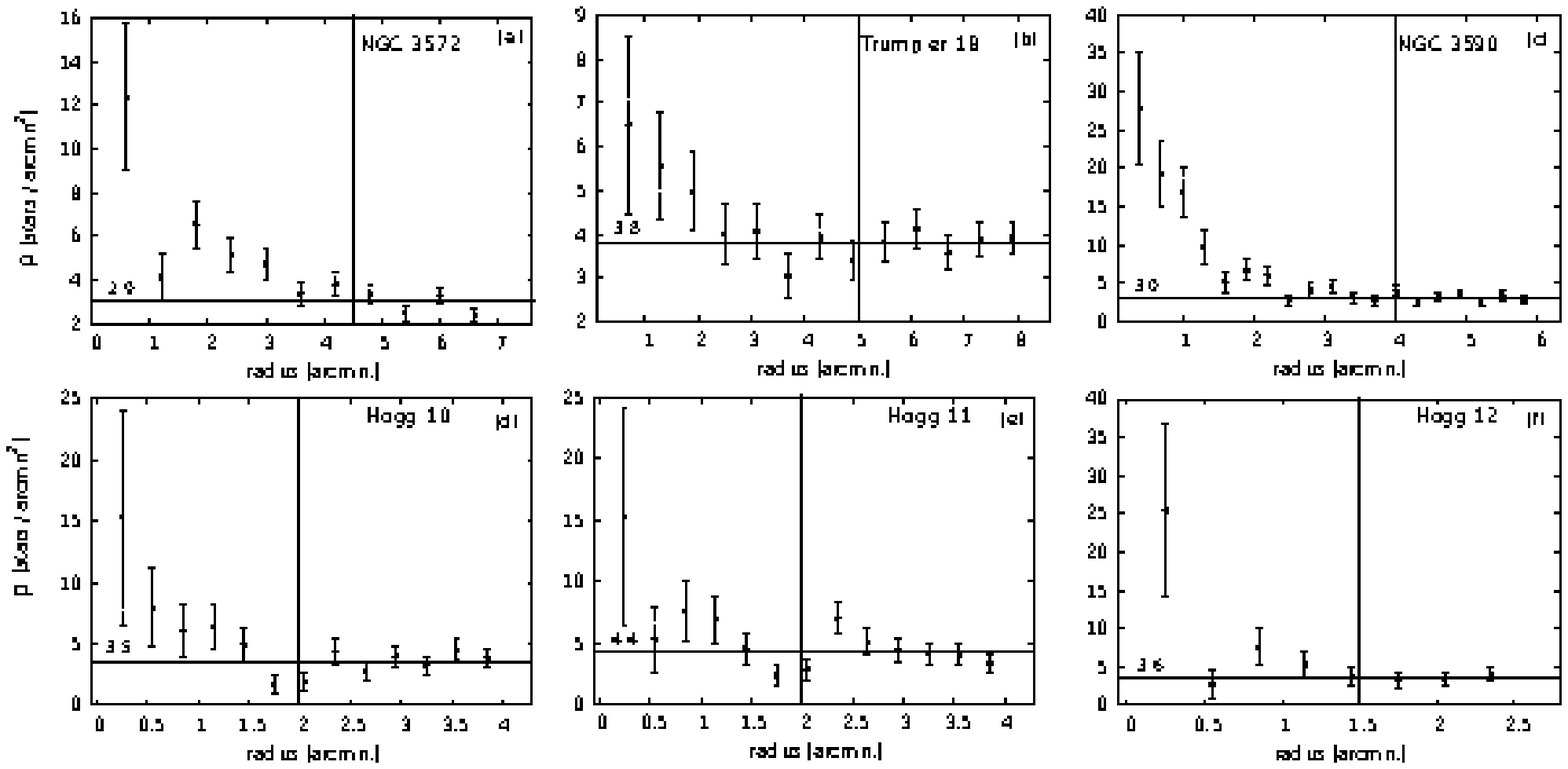}
\caption{RDPs obtained from brightest stars ($V < 16$) in each cluster region.  Vertical lines indicate the adopted cluster radii, while horizontal lines represent the mean level of the field.  $1\sigma$ Poisson error bars have been included.}
\label{fig:radial}
\end{center}
\end{figure*}


The geometrical structure of a star cluster (namely its center and radius) is an important indicator of its dynamical state. This is obvious for globular clusters, where the radial distribution of stars typically follows the well-known King profile (\citealt{1966AJ.....71..276K}). For an open cluster, the definition of both cluster center and radius is less obvious. The reason is that open clusters have irregular shapes in many cases, the number of stars is orders of magnitude smaller than in globulars, and the contamination of stars from the general Galactic field is significantly higher. Therefore the derivation of these quantities (namely where the center is, and where the cluster ends) inevitably includes
some degree of subjectivity.

It is, on the other hand, important to derive an estimate of the cluster center and radius, since this information, together
with the analysis of  cluster photometric diagrams, allows  us to derive fundamental parameters in a more reliable way.  The
reason is that stars within the cluster estimated radius define cleaner sequences in the photometric diagrams because the percentage of cluster members is higher. We stress that this is a crucial point. Detecting all cluster members would be a cumbersome task that would require acquiring kinematic multi-epoch information in addition to photometric data. This is beyond the scope of this work, which aims at deriving estimates of clusters fundamental parameters.  A widespread practice in the literature is to consider as the cluster border the distance from the adopted center where the cluster merges with the field, or, in other words, the radius at which the density of the cluster reaches the density of the field. We adopt this approach in the following.

To derive the cluster center, we adopted the following method. We started by using the center as provided by the catalog of
Dias et al. (2002) and constructed a radial density profile (RDP) using a set of rings around this center.  The ring size was chosen to be 10$\%$  of the radius reported by Dias et al. (2002). We initially adopted the radius of Dias et al. (2002) and then randomly changed these values in search for the coordinate pairs that produced the clearest central (inside the first circle) density peak. In this procedure, we explored a range of 30$^{\prime\prime}$ in both radial ascension and declination.

The cluster radius was then inferred using again the radial density profile and determining the point where the profile intersects the mean field level, as computed in a region away from the cluster. In some cases the density profile moves up and down the mean field level, causing more than one intersection. In these cases, we optimized the definition of the radius by looking at the resulting photometric diagrams, and then chose as radius the intersection of the profile with the mean field that produced the cleanest sequences in the photometric diagrams. The obtained RDPs are shown in Fig.~\ref{fig:radial}, where the field level is indicated with horizontal lines, while the adopted cluster radius is shown as vertical lines. The finally adopted cluster centers and radii values are reported in Table~\ref{tab:clusters}.

\subsubsection{Upper main-sequences of the clusters} \label{sec:upperms}

\begin{figure}
\begin{center}
\includegraphics[width=9cm]{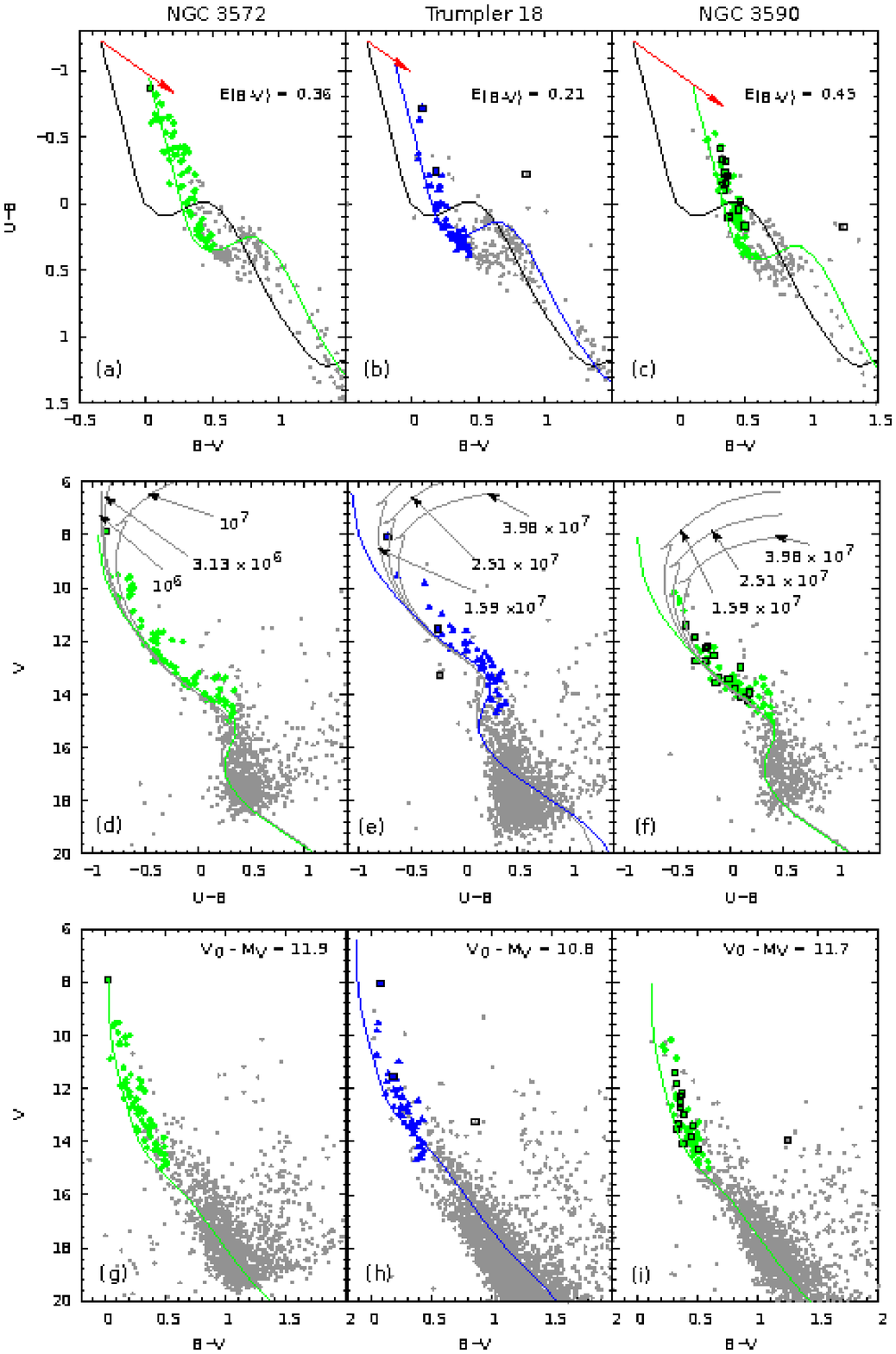}
\caption{TCDs and CMDs for stars located in the areas of NGC~3572, Trumpler~18, and NGC~3590. In all panels filled (green or blue) circles depict adopted likely members for each cluster; open black squares are stars with available spectral classification and small light gray dots are not classified or field stars. In the upper panels (a,b,c) we present the TCDs. Solid lines in them are ZAMSs (\citealt{2013JKAS...46..103S}), both in their normal location and shifted along the reddening path (red arrow) in agreement with the adopted color excesses (see Table~\ref{tab:clusters}).  In the middle panels (d,e,f) the gray lines represent solar metallicity isochrones, which include mass loss and overshooting (\citealt{2008A&A...482..883M}), for three different ages and
are plotted in years. In the lower panels (g,h,i), the solid lines are ZAMSs for each cluster, corrected for distance and reddening.}
\label{fig:phot1}
\end{center}
\end{figure}

\begin{figure}
\begin{center}
\includegraphics[width=9cm]{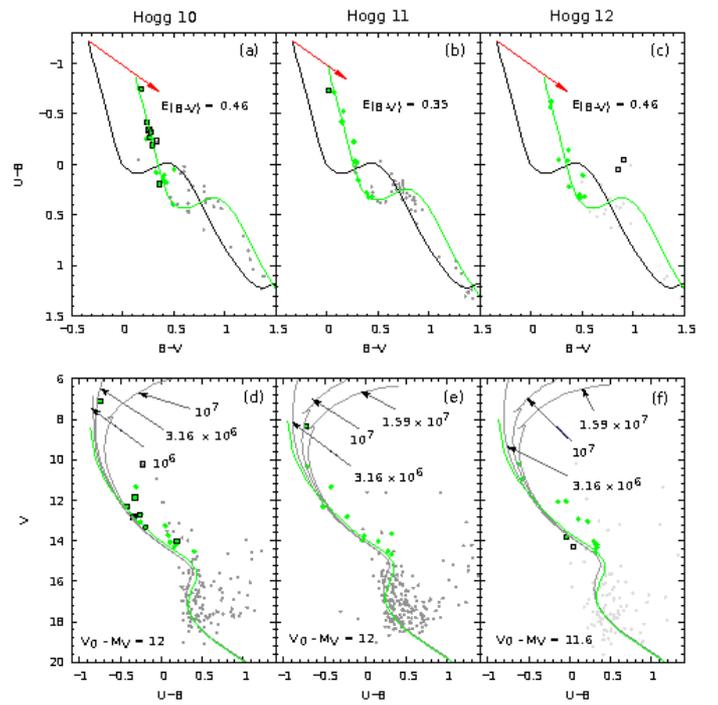}
\caption{Optical TCDs and CMDs of stars in the areas of Hogg~10, 11, and 12. Symbols and lines are the same as in Fig.~\ref{fig:phot1}. Numbers included are isochrone ages in years.}
\label{fig:phot3}
\end{center}
\end{figure}

\begin{figure}
\begin{center}
\includegraphics[width=9cm]{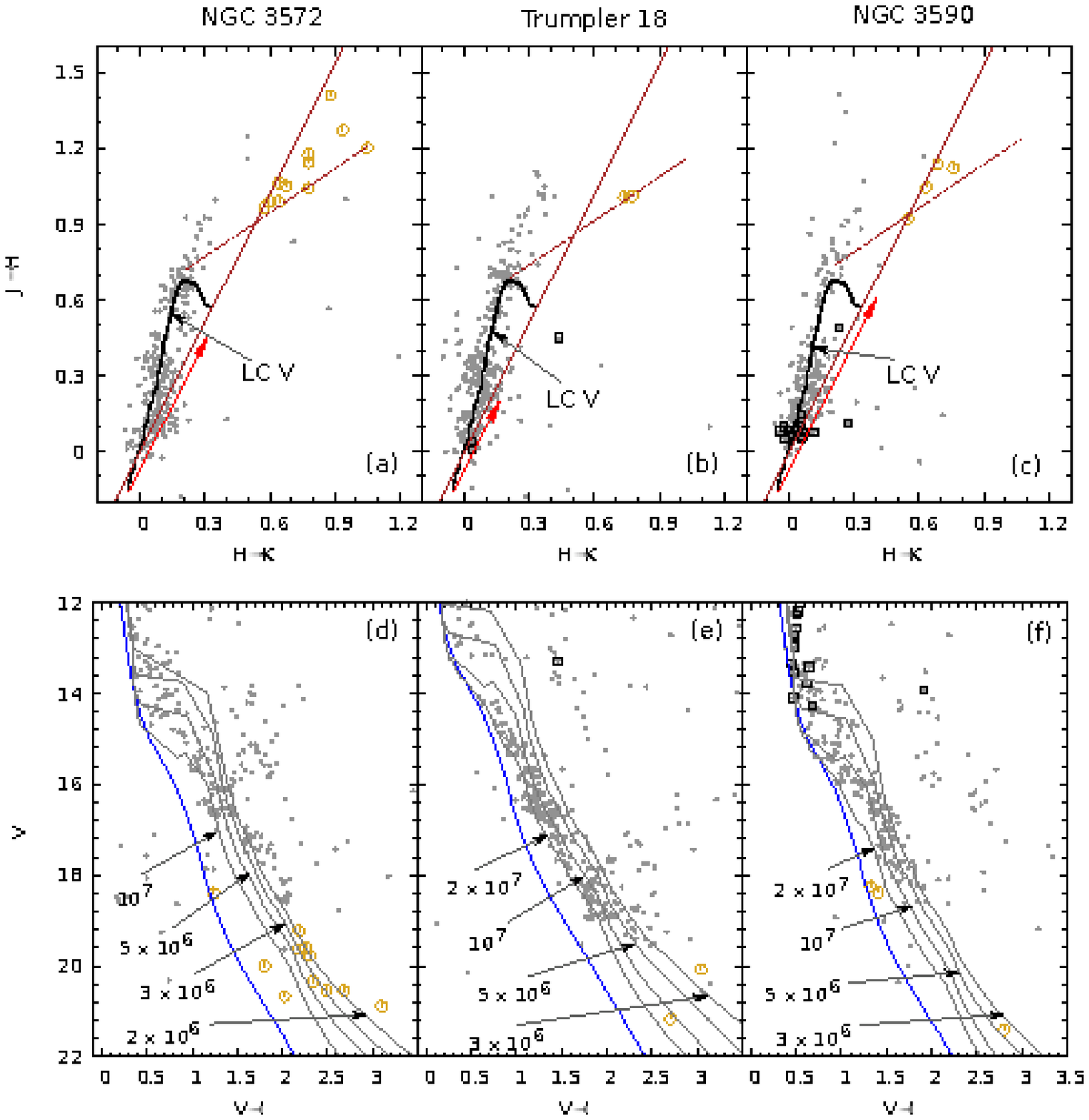}
\caption{Infrared photometric diagrams of stars located in the areas of NGC~3572, Trumpler~18, and NGC~3590.  In the upper panels (a,b,c), black lines show the intrinsic colors for MS stars according to (\citealt{1983A&A...128...84K}), arrows indicate the reddening paths, and lines indicate the limits adopted for the selection of YSO candidates (yellow open circles - see Sect.~\ref{sec:lowerms}). In the lower panels (d,e,f), we present decontaminated CMDs for each of the clusters (see Sect.~\ref{sec:lowerms}). Gray lines are PMS isochrones (\citealt{2000A&A...358..593S}) for four different ages in years.}
\label{fig:phot2}
\end{center}
\end{figure}

To obtain distances and reddening estimates for each cluster from their photometric data, we constructed their two-color diagrams (TCDs) and color-magnitude diagrams (CMDs), which are shown in Figs.~\ref{fig:phot1}-\ref{fig:phot2}. In the initial analysis we only took the brightest stars into account to avoid field contamination because it becomes more severe at fainter magnitudes.

To estimate the parameters for each cluster, we first shifted the zero-age main sequence (ZAMS; \citealt{2013JKAS...46..103S}) on the $U-B$ vs. $B-V$ diagrams according to the reddening law $E_{U-B}/E_{B-V} =
0.72 + 0.05 \cdot E_{B-V}$ until it was fit as an envelope for the bluest stellar distribution. After determining the $E_{B-V}$ values of the clusters in this way, we then fit a reddened ZAMS or main sequence (MS) to their CMDs to obtain their distance modulus ($V_{\circ} - M_V$), for which we adopted $R_V = 3.1$. We note that to detect possible deviations from the normal reddening law we used the cluster's $B-V$ vs. $V-I$ diagrams. These CMDs suggested a normal
behavior ($R_V = 3.1$) for all cluster populations, except for the most reddened and distant stars (see Sect.~\ref{sec:populations}).

Using the upper MS ($V > 15$) of the photometric diagrams, we adopted as cluster members those stars located between two envelope lines drawn around each ZAMS. These two envelopes were defined by shifting the ZAMSs by (or equivalent values): $\Delta V = 1$ above, and $\Delta (B-V) = 0.20-0.23$ to the right (envelope 1); and $\Delta V = 0.1$ down, and $\Delta (B-V) = 0.10-0.12$ to the right (envelope 2).

To estimate the age of the clusters, we also compared the observed upper MS of each cluster with theoretical isochrones (see Fig.~\ref{fig:phot1}). To this end, we used the set of isochrones given by \cite{2008A&A...482..883M} for solar metallicity, which consider mass loss and overshooting. Some scatter is present in our data, probably due to the presence of binaries, fast rotators, and/or differential reddening. This prevented a unique age solution, but it was still possible to obtain an estimate of those identified as$~\text{nuclear
age}$ values.


\subsubsection{Lower main-sequences of the clusters} \label{sec:lowerms}

The study of the lower MS in young open clusters allows searching for stellar formation signatures and/or candidates for young stellar objects (YSOs). This search involves the identification of sources with IR excess, which can be explained by emission from circumstellar dust. We assumed that these objects would be located in the lower part of the cluster's MS. These sources are expected to be low- and moderate-mass stars that were born from smaller gas lumps, which delayed their collapse to the ZAMS compared to the most massive and brightest objects.

To estimate the amount of stars in the lower MSs of the clusters, a statistical field decontamination of their CMDs was carried out. For this purpose, we constructed CMDs for appropriate comparison fields and looked for stars located in similar places on the cluster and field $V$ vs. $(B-V)$ and $V$ vs. $(V-I)$ diagrams, which were subtracted from the cluster CMDs (see \citealt{2003AJ....125..742G}). The resulting decontaminated cluster CMDs are presented in the lower panels of Fig.~\ref{fig:phot2}. A set of pre-main sequense (PMS) isochrones is included for comparison (\citealt{2000A&A...358..593S}), which allowed us to estimate the indicated as the$~\text{contracting
age}$ of the clusters.

Using available near-IR data ($JHK$, see Sect.~\ref{sec:otherdata}), we constructed the $(J-H)$ vs. $(H-K)$ diagrams shown in the upper panels of Fig.~\ref{fig:phot2}. They provide a good tool to distinguish between interstellar reddening and intrinsic color excess produced by circumstellar dust. To select probable YSOs on each cluster region, we
used the relation for classical T Tauri stars (CTTS) given by \cite{1997AJ....114..288M}: $(J-H)_{CTTS} = 0.58 \pm 0.11 \cdot (H-K)_{CTTS} +0.52 \pm 0.06$.

\section{Individual cluster properties} \label{sec:clusters}

The six clusters included in the studied region present several early-type stars (see Sect.~\ref{sec:intro}).  All these clusters are affected by relevant differential reddening that is characteristic of stellar formation regions. In this section we detail some of their particularities and summarize their adopted parameters in Table~\ref{tab:clusters}.

\subsection{NGC~3572 = C1108--599} \label{sec:ngc3572}

Previous studies recognized NGC~3572 as a young open cluster with a radius of about $5\farcm0$ (\citealt{2002A&A...389..871D}). It is considered to be the probable nucleus of a scattered group of OB stars
located in its vicinity, identified as Collinder~240 (\citealt{1976AJ.....81..155C}).

Our data show that this cluster has a typical RDP with a high central concentration (see Fig.~\ref{fig:radial}a). The radius we derived is consistent with previous estimates and defines a region that probably includes most cluster members. Near the center of the cluster lies a planetary nebula (PN G290.7+00.2; \citealt{2003A&A...408.1029K}) that has not been studied in detail
so far.

The brightest star in the cluster region (HD~97166) is the only object for which a spectral classification is available. It was recently identified as an O7.5IV + O9III binary system (\citealt{2014ApJS..211...10S}). This allowed us to obtain its
spectrophotometric parameters, yielding a distance modulus of $V_{\circ}-M_V = 12.5$ and a color excess of $E_{B-V} = 0.27$ (see Table~\ref{tab:spectra}).

The photometric diagrams for stars located in the cluster region (see Fig.~\ref{fig:phot1}adg) show that NGC~3572 has a well-defined and well-populated upper MS (green symbols), with a clear dispersion in color. These features allowed us to independently obtain (see Sect.~\ref{sec:upperms}) the cluster distance, color excess, and age. The values derived in this way are consistent with those estimated spectrophotometrically.

Our statistical analysis of the lower MS (see Sect.~\ref{sec:lowerms}) shows a clear overdensity in the decontaminated CMDs (see Fig.~\ref{fig:phot2}d), located roughly 2 magnitudes above the ZAMS. Additionally, our IR study revealed a significant amount of YSO candidates that appear in a consistent place in the CMDs (yellow open symbols). This evidence suggests that NGC~3572 has a large PMS population. Based on the location of this population in the $V$
vs. $V-I$ diagram, we derived an estimate of the cluster contraction age, which is similar to its nuclear age, suggesting a coeval stellar formation process.

\subsection{Trumpler~18 = C1109--604} \label{sec:tr18}

Our RDP (see Fig.~\ref{fig:radial}b) gives a radius of $5\farcm0$ for this cluster, in agreement with previous studies
(\citealt{2002A&A...389..871D}), and indicates that it is a scattered low-density object. Its basic parameters reveal that compared to NGC~3572, it is closer, has a lower reddening, and is more evolved.

The brightest cluster star (HD~97434) is classified as O7.5III(n)((f)). It has recently been recognized as a binary system (\citealt{2010RMxAC..38...30B}; \citealt{2014ApJS..211...10S}), but individual spectra are not available. Only one MS star of this cluster has a spectral classification (B7 V), from which we have obtained a distance modulus of $V_{\circ}-M_V = 11.0$ and a color excess of $E_{B-V} = 0.31$ (see Table~\ref{tab:spectra}).

Trumpler~18 also shows a probable PMS population whose contraction age is very similar to the nuclear age. This fact points to a probable scenario of coeval stellar formation; however, only a few objects were identified as YSO candidates by our IR analysis (see Fig.~\ref{fig:phot2}).

\subsection{NGC~3590 = C1110--605} \label{sec:ngc3590}

Previous studies recognized NGC~3590 as a typical young open cluster with a radius of about $3\farcm0$ (\citealt{2002A&A...389..871D}). Our RDP (see Fig.~\ref{fig:radial}c) indicates that it has a high stellar density core and an apparently extended halo. For this reason, we adopted a slightly larger radius ($4\farcm0$) for the cluster region, in an effort to include almost all its possible members.

The cluster region contains many B-type stars (14) (see Table~\ref{tab:spectra}), which allowed us to compute a mean spectrophotometric distance modulus of $V_{\circ}-M_V = 12.1 \pm 0.4$, and $E_{B-V}$ values between 0.45 and 0.57.


\begin{figure}
\begin{center}
\includegraphics[width=7cm]{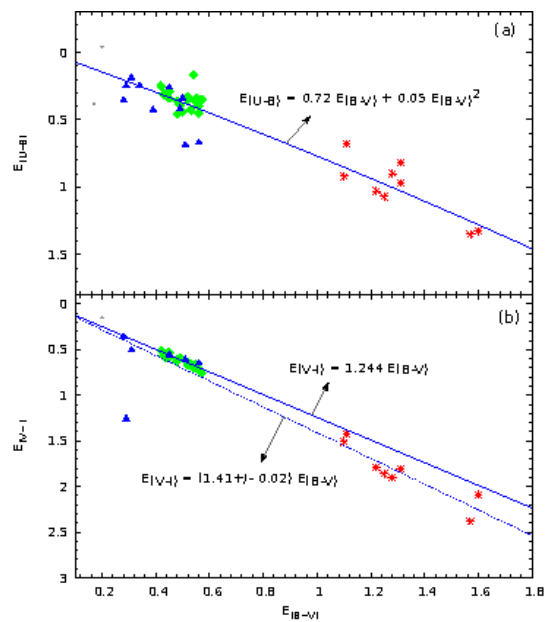}
\caption{$E_{U-B}$ vs. $E_{B-V}$, and $E_{V-I}$ vs. $E_{B-V}$ diagrams for stars with available spectral classification (see
  Table~\ref{tab:spectra}). Different colors indicate the newly identified $A$, $B,$ and $C$ stellar populations (see Sect.~\ref{sec:populations}). Solid lines are the normal relations ($E_{U-B} = 0.72 E_{B-V} + 0.05 E_{B-V}^2$ and $E_{B-V} = 1.242 E_{V-I}$), while the dotted line is a linear fit to to {\it Population C} ($E_{B-V} = 1.41 E_{V-I}$).}
\label{fig:excess_ratios-2}
\end{center}
\end{figure}


\begin{figure}
\begin{center}
\includegraphics[width=6cm]{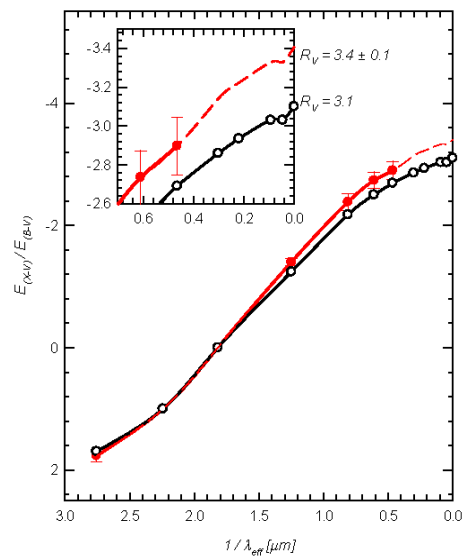}
\caption{Color difference method. The line that fits the black open circles represents the normal extinction behavior (\citealt{1968nim..book..167J} and \citealt{2002AJ....123.2559C}); while the line fitting the red filled circles shows the average extinction path obtained for the most reddened population ($C$). The red dashed line shows the extrapolation of the latter. Error bars are standard deviations.}
\label{fig:color_diff}
\end{center}
\end{figure}


As for NGC~3572, the photometric diagrams for NGC~3590 (see Fig.~\ref{fig:phot1}cfi) show that it has a well-populated upper MS (green symbols) with a clear dispersion. Hence we estimated its main parameters (distance, reddening, and age) in the usual way (see Sect.~\ref{sec:upperms}), obtaining values similar to those obtained spectrophotometrically. Its basic parameters reveal that NGC~3590 lies at a similar distance than NGC~3572, but shows higher reddening. It is also a more evolved object than NGC~3572. This might suggest a large-scale triggered cluster formation
situation for both objects.

We also identified a probable PMS population and YSOs candidates in the lower MS of this cluster (see Fig.~\ref{fig:phot2}) and estimated the corresponding contraction age. In this case, however, we found a more significant difference with the nuclear age. This might be interpreted as an indication that non-coeval stellar formation has occurred. On the other hand, a comparison of Figs.~\ref{fig:phot2}b and c suggests that this stellar population might be contaminated
by coronal and lower mass stars of Trumpler~18 in the NGC~3590 field.


\begin{figure*}
\begin{center}
\includegraphics[width=16cm]{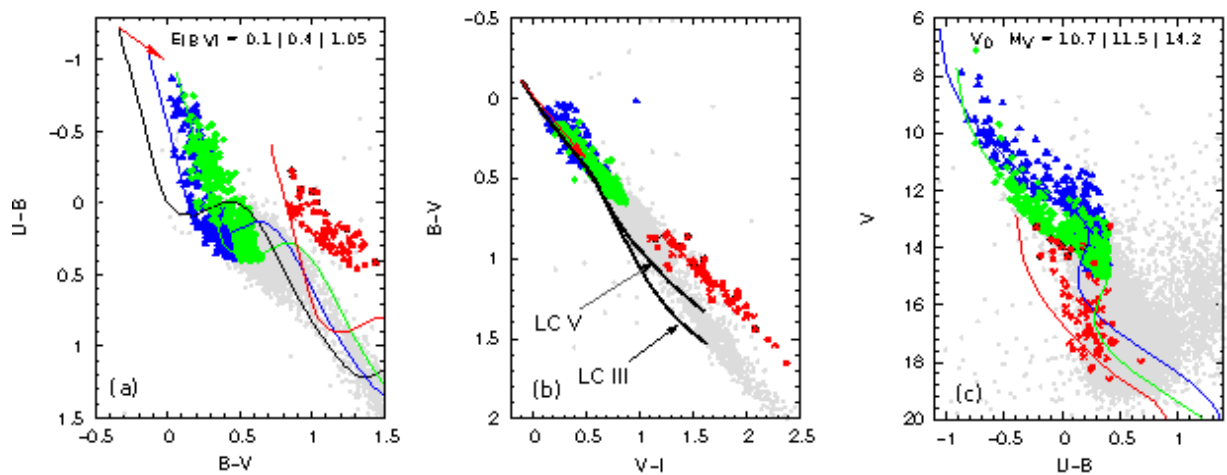}
\caption{Optical photometric diagrams for all stars in the region studied. Lines are the same as those in Fig.~\ref{fig:phot1}. The three young stellar populations we identified (see Sect.~\ref{sec:populations} for details) have been depicted with different colors: $A$ (blue), $B$ (green), and $C$ (red). Small light gray dots are unclassified stars. Black empty circles indicate the ten selected targets for the spectroscopic analysis (see Sect.~\ref{sec:populations}).}
\label{fig:phot1_all}
\end{center}
\end{figure*}


\subsection{Hogg~10, 11, and 12} \label{sec:hoggs}

These three clusters have similar angular sizes of about $1\farcm5 - 2\farcm0$. Although they contain a small number of bright stars, they exhibit typical open cluster RDPs (see Fig.~\ref{fig:radial}def).

Hogg~10 has several stars with spectral classification, from which we obtained a mean spectrophotometric distance modulus of $V_{\circ}-M_V = 12.8 \pm 0.4$, and $E_{B-V}$ values in the between 0.42 and 0.51 (see Table~\ref{tab:spectra}). Hogg~11
has only one star with a spectral classification that is identified as a giant. There is none for Hogg~12. It was therefore more difficult to obtain their basic parameters using only their photometric diagrams (see Fig.~\ref{fig:phot3}). We nonetheless provide the estimated values in Table~\ref{tab:clusters}.

We found that these three clusters have similar distances that are also similar to those derived for NGC~3572 and NGC~3590. This seems to agree with a scenario where all these objects belong to a common stellar population (see also Sect.~\ref{sec:populations}). Hogg~12 in particular is very close to NGC~3590, and they may be a binary system (as suggested by \citealt{2010PASP..122..516P}).

\subsection{Stellar masses and clusters mass function} \label{sec:masses}

\begin{figure}
\begin{center}
\includegraphics[width=7cm]{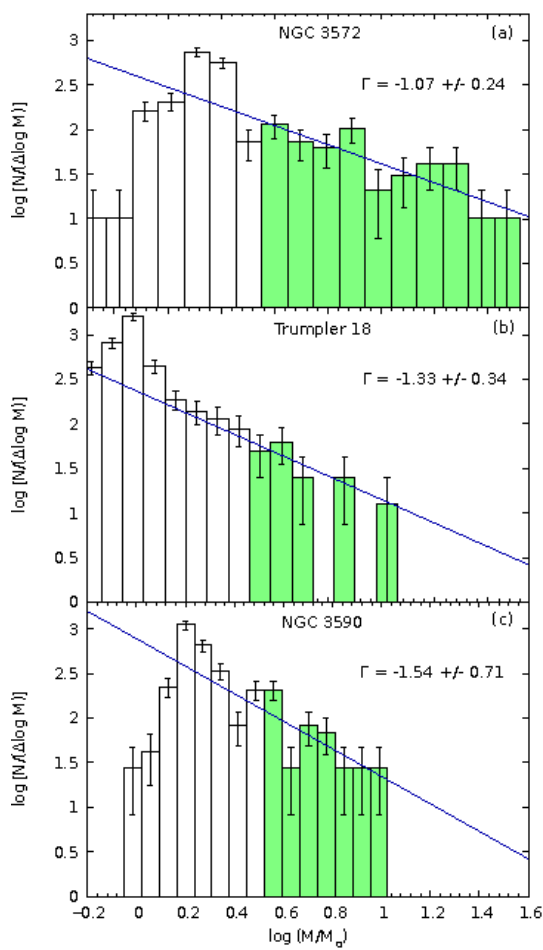}
\caption{IMFs for our three most conspicuous clusters; NGC~3572, Trumpler~18, and NGC~3590. Filled green bins ($M > 3 M_{\sun}$) indicate the values that were used to compute the power-law fits (see Table~\ref{tab:clusters}).  Error bars are Poisson uncertainties.}
\label{fig:imfs}
\end{center}
\end{figure}

We estimated stellar masses and their distribution for the three most prominent clusters: NGC~3572, Trumpler~18, and
NGC~3590. Stellar masses were computed by interpolating the best-fit isochrones for each cluster, and the corresponding histograms are presented in Fig.~\ref{fig:imfs}. Then, using only the bins corresponding to the high-mass range ($M > 3 M_{\sun}$), where the IMF can be modeled as a power law, we performed linear fits to the data and expressed them in the form: \\

$\log\left(\frac{N}{\Delta(\log m)}\right) = \Gamma \cdot  \Delta(\log m).$ \\

\noindent
The computed slope values ($\Gamma$) for each studied cluster is also indicated in Fig. \ref{fig:imfs} and Table~\ref{tab:clusters}.

To estimate the errors in the previous procedure, we repeated it for different distance values ($V_0 - M_V \pm 0.2$) for each cluster. We then obtained a $\delta M / M < 20\%$ and changes in the corresponding slopes lower than the indicated fitted errors.

In Table~\ref{tab:IMFs} we also report a list of IMF slopes for several star clusters in Carina, culled from the literature, and all on the same scale as the slopes derived in this work, namely on the scale where the Salpeter slope is -1.35. Our derived slope values and reading through the slope values in Table~\ref{tab:IMFs}
shows that regardless of the data quality and the method employed to build the mass distribution, the IMF slopes in this region are very uniform and close to the canonical Salpeter value. The only exception is NGC~3603, a massive starburst cluster, which shows a flatter mass distribution. This favors the idea of an universal mass distribution, similar to the Salpeter one.


\begin{table*}
\caption{Ages and IMF slope values for several open clusters in Carina.}
\centering
\begin{tabular}{lccccl}
\hline
Cluster & $\alpha_{J2000}$  & $\delta_{J2000}$ & Age [Myr] & $\Gamma$ & Reference \\
\hline
NGC~3293        & 10:35:51 & -58:13:48 &     $8 \pm 1$ & $-1.2  \pm 0.2$  & \cite{2003AA...402..549B}  \\
Trumpler~14     & 10:43:56 & -59:33:00 & $1.5 \pm 0.5$ & $-1.06 \pm 0.1$  & \cite{1996AAS..116...75V}  \\
Trumpler~16     & 10:45:10 & -59:43:00 &       $1 - 3$ & $-1.3  \pm 0.1$  & \cite{2012AJ....143...41H} \\
NGC~3532        & 11:05:39 & -58:45:12 &    $\sim 300$ & $-1.39 \pm 0.14$ & \cite{2011AJ....141..115C} \\
NGC~3603        & 11:15:07 & -61:15:36 &       $1 - 2$ & $-0.88 \pm 0.15$ & \cite{2013ApJ...764...73P} \\
IC~2944 (gr~a)  & 11:38:20 & -63:22:22 &      $\sim 3$ & $-1.00 \pm 0.35$ & \cite{2014MNRAS.443..411B} \\
IC~2944 (gr~b)  & 11:38:20 & -63:22:22 &      $\sim 3$ & $-1.02 \pm 0.37$ & \cite{2014MNRAS.443..411B} \\
Stock~16        & 13:19:29 & -62:38:00 &     $5 - 6.4$ & $-1.3  \pm 0.4$  & \cite{2005AA...430..471V}  \\
Collinder~272   & 13:30:26 & -61:19:00 &     $3 \pm 1$ & $-1.81 \pm 0.18$ & \cite{1997AAS..124...13V}  \\
Trumpler~21     & 13:32:14 & -62:48:00 &     $25 - 30$ & $-1.44 \pm 0.08$ & \cite{2001RMxAA..37...15G} \\
Lynga~1         & 14:00:02 & -62:09:00 &   $100 - 125$ & $-1.70 \pm 0.35$ & \cite{2003RMxAA..39...89V} \\
NGC~5606        & 14:27:47 & -59:37:54 &   $6.3 - 7.0$ & $-1.09 \pm 0.10$ & \cite{1994AAS..106..339V}  \\
NGC~6231        & 16:54:10 & -41:49:30 &       $3 - 5$ & $-1.14 \pm 0.10$ & \cite{1999AAS..137..233B}  \\
Havlen-Moffat~1 & 17:18:54 & -38:49:00 &       $2 - 4$ & $-0.65 \pm 0.05$ & \cite{2001AA...371..908V}  \\
\hline
\label{tab:IMFs}
\end{tabular}
\end{table*}


\section{Discussion}

\subsection{Stellar populations} \label{sec:populations}

We carried out a global analysis including all photometric and spectroscopic data available for the region. From the spectrophotometric analysis described in Sect.~\ref{sec:spec}, we obtained the relations $E_{U-B}$ vs. $E_{B-V}$, and $E_{V-I}$ vs. $E_{B-V}$, shown in Fig.~\ref{fig:excess_ratios-2}.  A fit to the data in the $E_{U-B}$ vs. $E_{B-V}$ seems to indicate that the normal reddening ratio is valid for all the stars, but the $E_{V-I}$ vs. $E_{B-V}$ diagram shows that the normal law is valid only for the less reddened stars. Because the most reddened stars clearly follow an abnormal law, we applied two different methods to them to estimate the selective absorption coefficient ($R_V = A_V/E_{B-V}$): the {\it \textup{excess ratio method}} and the {\it \textup{color difference method}} (see \citealt{1994AAS..106..339V} for details). In the first method, we simply performed a linear fit, obtaining a slope of $1.41 \pm 0.02$ compatible with $R_V = 3.5 \pm 0.05$ (see Fig.~\ref{fig:excess_ratios-2}). To apply the second method, we used the multiband information available for each star (see Fig.~\ref{fig:color_diff}). This information includes our $UBVI_{KC}+JHK$ catalog together with the Wide-field Infrared Survey Explorer ($WISE$\footnote{http://wise2.ipac.caltech.edu/docs/release/allsky/}; \citealt{2010AJ....140.1868W}) data ($W_1$, $W_2$, $W_3$, $W_4$ bands). We obtained $R_V = 3.4 \pm 0.1$.

In Fig.~\ref{fig:phot1_all} we present the photometric diagrams for all the objects in our catalog. Based on the color excess ratios derived for the individual clusters in Sect.~\ref{sec:clusters}, these diagrams allowed us to separate part of the objects in three young stellar populations: $A$, $B$ and $C$, respectively depicted in blue, green, and red. The stellar distributions in the CMD and TCDs suggest that the differences among the groups might be interpreted in terms of distance and different color excess ranges. Hence, using the procedure described in Sect.~\ref{sec:upperms}, it was possible to obtain approximate basic parameters for each population, which are given in Table~\ref{tab:populations}.


\begin{table}
\begin{center}
\caption{Parameters of the three young stellar populations}
\label{tab:populations}
\fontsize{9} {14pt}\selectfont
\begin{tabular}{cccc}
\hline
Population & $E_{B-V}$   & $R_V$ & $V_{\circ}-M_V$   \\
\hline
 A         & 0.10 - 0.40 & 3.1   & 10.7 - 11.8 \\
 B         & 0.30 - 0.60 & 3.1   & 11.5 - 12.6 \\
 C         & 0.95 - 2.05 & 3.5   & 14.1 - 15.2 \\
\hline
\label{tab:groups}
\end{tabular}
\end{center}
\end{table}


\begin{figure}
\begin{center}
\includegraphics[width=8cm]{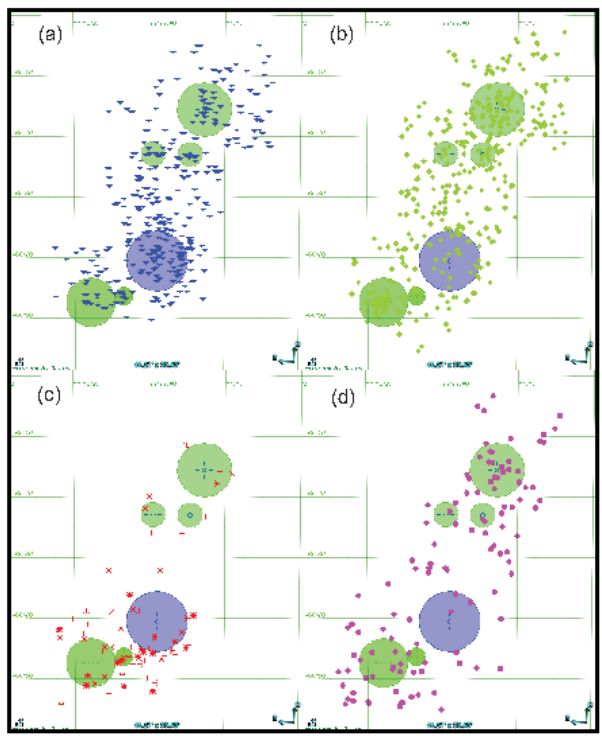}
\caption{ Spatial distribution of different stellar populations (see Sect.\ref{sec:populations}) in the covered area: a) $Population~A$, b) $Population~B$, c) $Population~C$, and d) YSO candidates. Large circles have the same meaning as in Fig,~\ref{fig:ccds}.}
\label{fig:populations}
\end{center}
\end{figure}


Regarding the spatial distribution of these populations (see Fig.~\ref{fig:populations}), the nearest one ($A$) is concentrated around Trumpler~18 at a distance of 1.4-2.3~kpc, the second nearest ($B$) is spread almost uniformly throughout the entire cluster region (represented by the other five open clusters) at a distance of 2.0-3.3~kpc, and the farthest one ($C$) is located preferentially toward the southern part of the field, with a clear concentration between Hogg~12 and Trumpler~18.  From this last group, we chose ten objects (see Fig. \ref{fig:phot1_all}, black empty circles). Based on spectroscopic analysis (see Sect.~\ref{sec:spec_class}), we found that these selected targets are OB-stars and are scattered between 6.6 and 11 kpc. This spatial distribution, together with the parameters found for these three populations, suggest that Trumpler~18 is a concentration of {\it Population A} objects, and that all other clusters belong to the {\it Population B}. Consistently with its high reddening, {\it Population C} lies far behind the other two. We note that the three groups of young stars found in this work present a scenario similar to that revealed from other studies on closer Galactic directions (see, e.g., \citealt{2009A&A...493...71C}, \citealt{2014MNRAS.443..411B} or \citealt{2015MNRAS.450.3855M}).

\begin{figure}
\begin{center}
\includegraphics[width=9cm]{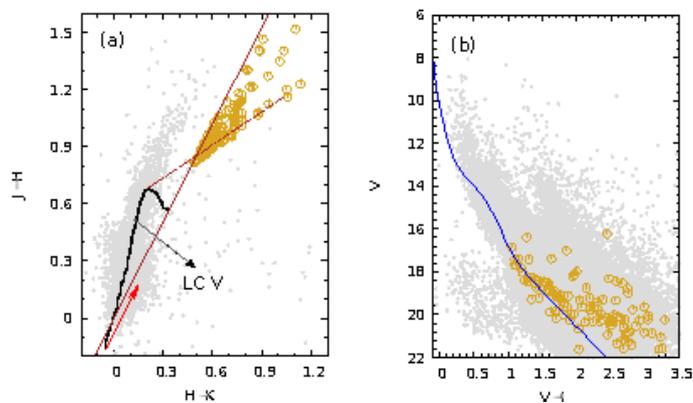}
\caption{Infrared photometric diagrams for all stars in the region. Symbols are the same as in Fig.~\ref{fig:phot2}.}
\label{fig:phot2_all}
\end{center}
\end{figure}

The IR photometric diagrams for all stars in the region studied shown in Fig.~\ref{fig:phot2_all} reveal several objects with an apparent IR excess (see Sect.~\ref{sec:lowerms}). From their spatial distribution we could infer that they seem to be concentrated in two groups, one toward the north and other toward the south, apparently associated with the main open clusters in the region. We therefore interpret these objects as the result of a mass segregation effect in the fainter stellar population of each cluster. This effect could have been caused by a dynamic process. However, the youth of most of the studied clusters may mean that it might reveal a primordial spatial mass distribution, as previously discussed.

The group of clusters we studied is located SE from the core of the Carina nebula (e.g., Trumpler~16), and at a higher Galactic longitude and lower Galactic latitude, namely closer to the formal Galactic plane (see Fig.~\ref{fig:ccds}). NGC~3572 is the closest to Trumpler~16, while NGC~3590 is the most distant. Between NGC~3572 to NGC~3590 there seems to be an age gradient, with the youngest objects being located toward the Carina core. This age gradient is also seen from the distribution of the gas and nebulosity, which is dominant in the northwest region. We tend to exclude, however, any possible relation in terms of star formation with the great Carina nebula, since the distance is too large, $\sim 3^{\circ}$. Another interesting piece of information is that the NW younger population is generally closer to us than the SE older population. It would be extremely interesting to extend the present study to the region $287^o \leq l \leq 290^o$ to verify whether this trend continues.


\subsection{Spiral structure toward Carina}

\begin{figure}
\begin{center}
\includegraphics[width=8cm]{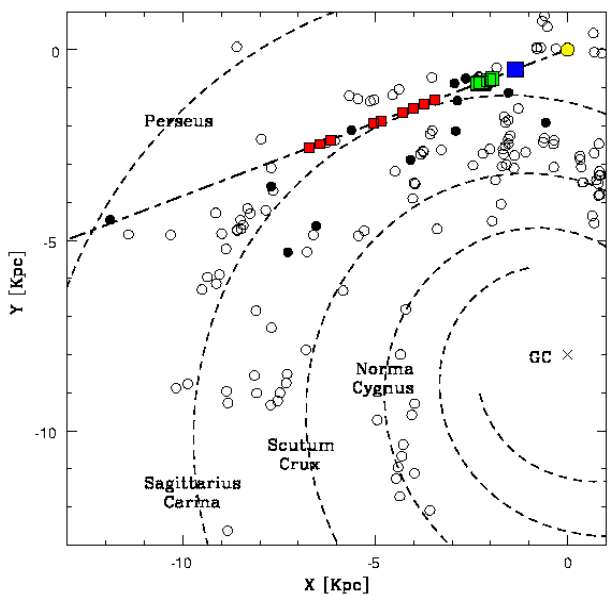}
\caption{Spatial relation between the studied Galactic direction (dashed line) and the \cite{2008AJ....135.1301V} model with four arms. A distance to the Galactic center of 8.0 kpc was assumed. The yellow circle indicates the position of the Sun. Colored squares depict the three stellar populations discovered in this work (see Sect.~\ref{sec:populations}). Other relevant stellar populations have been plotted as a reference: filled circles represent open clusters and young associations with optical $UBVI$ data from several studies, and open circles indicate HII tracers (\citealt{2009A&A...499..473H}).}
\label{fig:galactic}
\end{center}
\end{figure}

In Fig.~\ref{fig:galactic} we show the spatial distribution of the various stellar groups we identified and characterized in this study. This is the distribution projected in the plane of the Milky Way as seen from the North Galactic Cap. For better visualisation, the whole fourth quadrant ($270^o \leq l \leq 360^o$) is shown. We complement the result of the present study with data from previous papers (see, e.g., \citealt{2014MNRAS.443..411B}), such as young open clusters and HII regions, to provide a more comprehensive picture of the fourth quadrant. The figure also includes the four-arm Galactic model from Valle\'e \cite{2008AJ....135.1301V}. We recall that  this model is the result of a statistical interpolation of many different tracers (CO clouds, HII regions, young star clusters, HI, and masers) and is by no means intended to provide a solid and complete description of the Milky Way spiral structure. In this sense, the accumulation of more data with more reliable distance estimates is extremely valuable to improve the model predictions. This is evident from Fig.~\ref{fig:galactic}, where  data are in general much more scattered than the precise logarithmic line representing the arm. This holds even taking the typical width of a spiral arm into account, which in the inner disk is around 1 kpc (\citealt{2000A&A...358..521B}). The direction to the Galactic center is particularly indicative in this respect, and previous studies have emphasized that the distribution of the young stellar population does not seem to be discrete toward the bulge, but continuous, and Fig.~\ref{fig:galactic} confirms this scenario.

The situation toward the Carina-Sagittarius arm seems to be less complex. We confirm previous findings that young groups are present in the direction at discrete distances (\citealt{2009A&A...493...71C}, \citealt{2012Ap&SS.337..303T}). Inspecting Fig.~\ref{fig:galactic} closely, we can suggest the following:

\begin{itemize}
\item  Trumpler~18 is offset with respect to the expected location of the Carina-Sagittarius arm. This can be explained by the age difference to the other studied groups (Trumper 18 is older). The cluster might have had the time to drift away from its birthplace, close to the arm.
\item The $B$ group  is consistent with Valle\'e predictions taking the arm width into account.
\item Stars belonging to the $C$ group are quite scattered in distance and do not always follow the arm. They depart from the arm at larger distances. This might indicate that the Carina-Sagittarius arm in this region has a larger pitch angle than suggested by Valle\'e.
\end{itemize}


\section{Conclusions} \label{sec:conclusion}

We have presented a deep and homogeneous optical ($UBVI_{KC}$) photometric catalog of a wide region of the Galactic plane at $l~=~291^{\circ}$, which was cross-correlated with near-IR data from $2MASS$. We also presented a spectral classification for several conspicuous stars in the area. Our spectroscopy, together with published spectroscopic classifications of other objects of interest in the field, allowed for a reliable analysis of the stellar energy distribution, helping prevent possible degeneration in the photometric diagrams.

We estimated the main parameters (distance, size, reddening, age, and mass distribution) of the six open clusters in the region. We applied our numerical code that allowed verifying the obtained results jointly and not just individually.
Our estimated values for the radii, although they should be considered as upper limits, are equal to or smaller than those provided by \cite{2002A&A...389..871D} in five of the six star clusters.

We found that the main clusters (NGC~3572, Tr~18, and NGC~3590) present signatures of a probable coeval PMS population and identified YSOs candidates that need to be confirmed with follow-up spectroscopy. Additionally, we detected three young stellar populations for which we derived approximate basic parameters. The nearest is concentrated around Trumpler~18. The second nearest is spread almost uniformly throughout the entire cluster region (represented by the other five open clusters). And finally, the farthest population is abnormally reddened and young, confirmed by our spectroscopic data, from which we identified two O7-type stars, eight B-type stars, and one Be star.

Finally, we interpreted the spatial distribution of the young population in terms of the spiral structure of the Milky Way and identified a promising relation between age, distance, and position along the arm of the various clusters, which deserves further investigation  by extending this type of study to regions closer to the core of the great Carina nebula.

\begin{acknowledgements}

AML, GB and RG acknowledges support from CONICET (PIPs 112-201101-00301 and 112-201201-00298). GB also acknowledges financial support from the ESO visitor program that allowed a visit to ESO premises in Chile, where part of this work was done.
EC acknowledges support by the Fondo Nacional de Investigaci\'on Cient\'{\i}fica y Tecnol\'ogica (project No. 1110100, Fondecyt) and the Chilean Centro de Excelencia en Astrof\'{\i}sica y Tecnolog\'{\i}as Afines (PFB 06).
This paper is based on
a) observations at Cerro Tololo Inter-American Observatory, National Optical Astronomy Observatory which is operated by the Association of Universities for Research in Astronomy (AURA) under a cooperative agreement with the National Science Foundation;
b) observations obtained at the Gemini Observatory, which is operated by the Association of Universities for Research in Astronomy, Inc., under a cooperative agreement with the NSF on behalf of the Gemini partnership: the National Science Foundation (United States), the National Research Council (Canada), CONICYT (Chile), the Australian Research Council (Australia), Minist\'{e}rio da Ci\^{e}ncia, Tecnologia e Inova\c{c}\~{a}o (Brazil) and Ministerio de Ciencia, Tecnolog\'{i}a e Innovaci\'{o}n Productiva (Argentina).  The authors are much obliged for the use of the NASA Astrophysics Data System, of
the $SIMBAD$ database and $ALADIN$ tools (Centre de Donn\'es Stellaires --- Strasbourg, France), and of the WEBDA open cluster database.
This publication also made use of data from:
a) the Two Micron All Sky Survey, which is a joint project of the University of Massachusetts and the Infrared Processing and Analysis Center/California Institute of Technology, funded by the National Aeronautics and Space Administration and the National Science Foundation;
b) the Wide-field Infrared Survey Explorer, which is a joint project of the University of California, Los Angeles, and the Jet Propulsion Laboratory/California Institute of Technology, funded by the National Aeronautics and Space Administration.  We thank R. Mart\'{\i}nez and H. Viturro for technical support. We acknowledge our referee for the helpful comments and constructive suggestions that helped to improve this paper.
\end{acknowledgements}

\bibliographystyle{aa} 
\bibliography{biblio} 




\end{document}